\documentclass{aa} 

\usepackage{multirow}
\usepackage{graphicx}
\usepackage{color}

\usepackage{txfonts}
\begin{document}

   \title{Gas and galaxies in filament between clusters of galaxies:}
   \subtitle{The study of A399-A401}

   \author{V. Bonjean\inst{1,2},
          N. Aghanim\inst{1},
          P. Salomé\inst{2},
          M. Douspis\inst{1},   
          \and       
          A. Beelen\inst{1}
          }

   \institute{Institut d'Astrophysique Spatiale, CNRS (UMR 8617) Université Paris-Sud, Bâtiment 121, Orsay, France
   \email{victor.bonjean@ias.u-psud.fr}
         \and
             LERMA, Observatoire de Paris, PSL Research University, CNRS, Sorbonne Universités, UPMC Univ. Paris 06, 75014, Paris, France
             }

   \date{Received XXX; accepted XXX}

% \abstract{}{}{}{}{} 
% 5 {} token are mandatory
 
  \abstract
   {
     We have performed a multi-wavelength analysis of two galaxy cluster systems selected with the thermal Sunyaev-Zel'dovich (tSZ) effect and composed of cluster pairs and an inter-cluster filament. We have focused on one pair of particular interest: A399-A401 at redshift $z\sim0.073$ seperated by 3 Mpc. We have also performed the first analysis of one lower significance newly associated pair: A21-PSZ2 G114.09-34.34 at $z\sim0.094$, separated by 4.2 Mpc. We have characterised the intra-cluster gas using the tSZ signal from Planck and, when this was possible, the galaxy optical and infra-red (IR) properties based on two photometric redshift catalogs: 2MPZ and WISExSCOS. From the tSZ data, we measured the gas pressure in the clusters and in the inter-cluster filaments. In the case of A399-A401, the results are in perfect agreement with previous studies and, using the temperature measured from the X-rays, we further estimate the gas density in the filament and find $n_{\mathrm{0}}=(4.3\pm0.7)\times10^{-4} \mathrm{cm}^{-3}$. The optical and IR colour-colour and colour-magnitude analyses of the galaxies selected in the cluster system, together with their Star Formation Rate, show no segregation between galaxy populations, in the clusters and in the filament of A399-A401. Galaxies are all passive, early type, and red and dead. The gas and galaxy properties of this system suggest that the whole system formed at the same time and corresponds to a pre-merger, with a cosmic filament gas heated by the collapse. For the other cluster system, the tSZ analysis was performed and the pressure in the clusters and in the inter-cluster filament was constrained. However, the limited or nonexistent optical and IR data prevent us from concluding on the presence of an actual cosmic filament or from proposing a scenario.
     }
   \maketitle   
\section{Introduction}\label{intro}
Cosmic structures formed through accretion of matter, from the small scales quantum fluctuations in the very early Universe to the tens of Mpc-scale filaments and super-clusters. This scenario was largely corroborated by cosmological
numerical simulations such as Millenium \citep{springel}, Horizon \citep{teyssier}, or Illustris \citep{vol}, reproducing the evolution of the 80\% dark matter (DM) content of the universe. Observations from large surveys (e.g. SDSS, \cite{tempel}; GAMA, \cite{alpaslan}) allow us to probe the large-scale distribution of galaxies that amount 4\% in the form of cold baryons (cold gas and stars in galaxies). The warm-hot and hot gas represents the remaining 16\% of matter. Both simulations and observation indicate that matter is distributed in the form of a cosmic web, an ensemble of interconnected filaments, sheets, and voids. The baryonic matter flows along the gravitational potential well of DM filaments into the connecting nodes of the cosmic web: the galaxy clusters, where the gas is virialised and heated up to high temperatures (of order $10^8$K).
\\
\\
As they are build up over time from mergers and interactions of smaller systems \citep[e.g.,][]{navarro,springel}, galaxy clusters are naturally connected to the cosmic web via the filaments. Strategies to probe the cosmic web are thus associated with our ability to probe filamentary structures between clusters or in their outskirts. This is in principle possible via observation of the galaxy distribution, the weak gravitational lensing, the X-ray emission from hot gas, and the thermal Sunyaev-Zel'dovich (tSZ) effect \citep{sz}, but the search for filaments linking the clusters to the cosmic web is hard. It has gained a lot of interest and focused mostly on two cases: filaments in the outskirts of individual clusters and inter-cluster filaments (or bridges) in pairs of clusters. In the first case, \cite{eckert} have detected large-scale structures of several Mpc in the outskirts of the galaxy cluster Abell 2744 at redshift $z=0.306$, combining observations of the X-ray emission, galaxy over-densities, and weak lensing. For the DAFT/FADA cluster sample \citep{guennou}, \cite{durret} searched for and found filaments in clusters' outskirts with 2D galaxy densities obtained with CFHT and SUBARU observations. In the second case i.e. cluster pairs, the inter-cluster filament or the bridge is expected to be denser with a hotter gas and thus, in principle, easier to detect in particular in the X-rays and tSZ effect \citep{dolag}. Cluster pairs are thus good targets and they have thus been the subject of numerous studies. The photometric properties of the galaxies in the inter-cluster filament, their star-formation evolution, their weak lensing properties, etc. were performed in many selected cluster (or group) pairs \citep[e.g.,][]{fadda,gallazzi,louise,zhang,martinez,epps}. Galaxy clusters may show substructures or evidence of dynamical effect: they merge, interact, and accrete smaller groups. The galaxy properties derived from optical and near-infrared data thus need to be combined, in multi-wavelength analysis, with the study of cluster gas content. The gas properties of cluster pairs were therefore also investigated mostly using X-ray observations. This is the case of the particular pair Abell 399 - Abell 401, thoroughly studied using data from ASCA, ROSAT, Suzaku and XMM-Newton \citep[e.g., ][]{kara,ulmer,fujita3,fabian,sakelliou,fujita,fujita2}.
\\
\\
Another probe of the diffuse ionised hot gas in the clusters and cluster pairs (more generally cluster systems) is the tSZ effect. Its amplitude measures the gas pressure, $P_{\mathrm{e}}$, and it is quantified by the Compton parameter $y$:
\begin{equation}
y=\frac{\sigma_{\mathrm{T}}}{m_{\mathrm{e}}c^{2}}\int P_{\mathrm{e}}(l)\mathrm{d}l=\frac{\sigma_{\mathrm{T}}}{m_{\mathrm{e}}c^{2}}\int n_{\mathrm{e}}(l)k_{\mathrm{b}}T_{\mathrm{e}}(l)\mathrm{d}l,
\end{equation}
where $\sigma_{\mathrm{T}}$ is the Thomson cross-section, $m_{\mathrm{e}}$ the mass of the electron, $c$ the speed of light, $k_{\mathrm{b}}$ the Boltzmann constant, and $n_{\mathrm{e}}(l)$ and $T_{\mathrm{e}}(l)$ respectively the density and the temperature of the free electrons along the line of sight. The $n_{\mathrm{e}}$ dependence of the tSZ effect, compared with the $n_{\mathrm{e}}^2$ dependence of the X-ray emission, makes the tSZ effect the good tracer of diffuse gas regions. The Planck survey \citep{planck1} provides tSZ maps over all the extragalactic sky \citep{szmap}. This has driven some recent studies on the inter-cluster filaments from tSZ data \cite[e.g., ][]{planck135}. 
\\
\\
%The expansion of the Universe has distributed the baryons issued of the primordial nucleosynthesis in several physical states all along the epochs. Today these baryons are mainly located in intergalactic hot and cold gas, and in gas into galaxies or stars. \cite{fukugita} have made a baryon census at low redshift and pointed out the lack of baryon fraction $\Omega_{\mathrm{b}}$ compare to the one derived by using the Cosmic Microwave Background (CMB) data and the large scale structures observations at high redshifts \citep{fukugita,planckcosmo}. Numerical simulations \citep{cen,dave} predict these missing baryons in a hot and dense phase, with a temperature $T=10^5-10^7$K, along the large-scale filaments of the cosmic web, and around galaxy clusters. These regions are the so-called Warm Hot Intergalactic Medium (WHIM). However, these gas reservoirs are very hard to detect with the current telescopes facilities. \textcolor{red}{I do not think that this is the right introduction to the paper. I think that you should rather introduce the structure formation in which clusters (as seen in sims) are formed at the intersection of the cosmic filaments. Therefore people are now interested in detecting these filaments in between or in the outskirts of the clusters}.
In this study, we focus on cluster pairs and we perform a multi-wavelength analysis with tSZ data from Planck and optical and near-infrared data from SDSS and WISE \citep{wise}. We derive the properties of both the gas and the galaxies in the inter-cluster filament to probe its possible origin. In Sect.~\ref{data}, we present the data used for the multi-wavelength study. We explain, in Sect.~\ref{selectioncluster}, our selection of the cluster pairs studied in this work. In Sect.~\ref{gas} and in Sect.~\ref{galaxies}, we focus on the pair A399-A401. In the former, we perform the analysis of the tSZ data and derive the extension of the filaments, its pressure and its density. In the latter, we explore the properties of the galaxies in the clusters A399 and A401, and in the inter-cluster filament. We discuss our results in Sect.~\ref{discussion} and we present our conclusions in Sect.~\ref{conclusion}. We adopt all through this study the Planck 2015 cosmological parameters of \cite{cosmo15}. In particular, all distances and scales are computed with $h=0.6774$.

\begin{figure*}[t]
\centering
\includegraphics[width=0.5\textwidth]{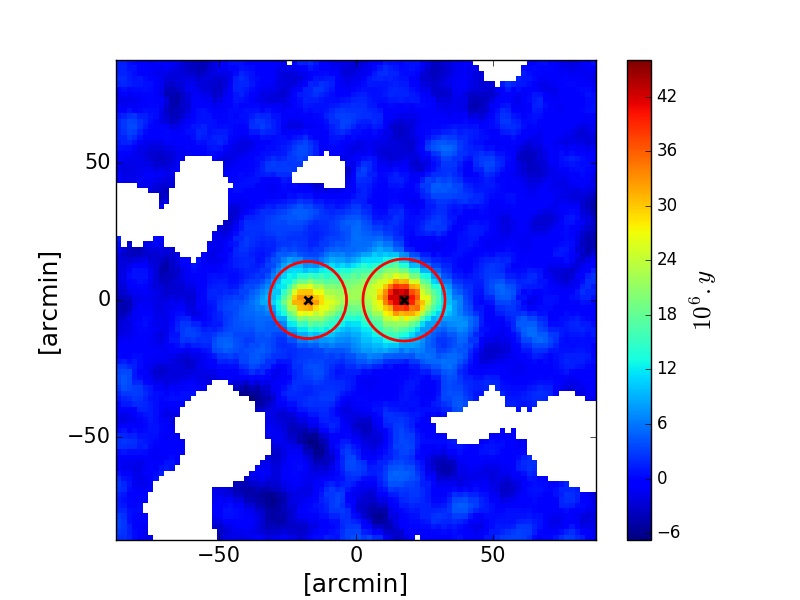}\includegraphics[width=0.5\textwidth]{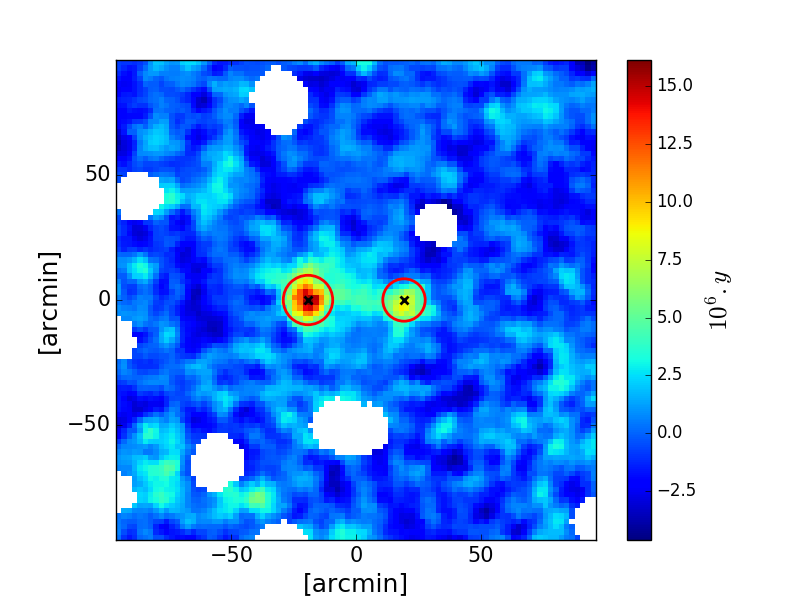}
\caption{\label{maps} Extracted masked patches of the Planck tSZ map of the two selected pairs. Left: the pair A399-A401. Right: the pair A21-PSZ2G 114.90-34.35. The red circles indicate radii $r_{500}$ of the clusters and the white pixels are masked from the Planck Catalog of Compact Sources and from other clusters in the fields. The black crosses show the centers of the tSZ clusters.}
\end{figure*}

\section{Data}\label{data}

\subsection{Meta-catalogs of galaxy clusters}

In order to identify systems of galaxy clusters with filaments, we use the publicly available SZ cluster database\footnote{\url{http://szcluster-db.ias.u-psud.fr/}}. It is the union of tSZ cluster catalogs obtained with different instruments: Planck, ACT, SPT, Carma, and AMI (see on the database's website references to the catalogues used). The catalog contains 2690 sources, and for 1681 sources which are identified as clusters, the redshifts and mass proxies $M_{500}$ (defined in \cite{mprox}) are available. The redshifts range is between $z_{\mathrm{min}}=0.011$ and $z_{\mathrm{max}}=1.7$ with a median value $z_{\mathrm{med}}\sim0.31$.

\subsection{Planck data}

Planck \citep{planck1} was the third satellite to measure the CMB temperature anisotropies. It observed the full-sky more than five times in nine frequencies from 30 to 857 GHz with two instruments: HFI (High Frequency Instrument, \cite{lamarre}; \cite{phfi}) and LFI (Low Frequency Instrument, \cite{bersanelli}; \cite{mennella}). LFI covers the 30, 44, and 70 GHz bands with 33.29, 27.00, and 13.21 arcmin angular resolution and HFI the 100, 143, 217, 353, 545 and 857 GHz bands with 9.68, 7.30, 5.02, 4.94, 4.83 and 4.64 arcmin angular resolution respectively.
\\
\\
The frequencies of Planck have been specifically chosen to measure the tSZ effect from galaxy clusters. Based on two component separation techniques, the Planck collaboration has constructed full-sky maps of the tSZ Compton parameter at a resolution of 10 arcmin, using the six highest frequencies \citep{szmap}. We checked that using either of the two maps does not alter our results. In the following, we thus use the Planck 2015 map from \citep{szmap} constructed with the Modified Internal Linear Combination Algorithm (MILCA) \citep{guillaume}. This map\footnote{The tSZ map is available at \url{http://pla.esac.esa.int/pla/}.} is in HEALPIX format \citep{healpix}, with $n_{\mathrm{side}}$=2048 and a pixel size $\theta_{\mathrm{pix}}$=1.7 arcmin. 

\subsection{WISE}

The WISE satellite \citep{wise} surveyed the whole sky at wavelengths 3.4, 4.6, 12 and 22$\mu$m, with angular resolutions respectively 6.1, 6.4, 6.5 and 12 arcsec. The All-WISE Source Catalog\footnote{Available at \url{http://wise2.ipac.caltech.edu/docs/release/allwise/expsup/sec1_3.html#src_cat}} contains positions, motion measurements, photometry and ancillary information for 747,634,026 sources detected on AllWISE Atlas Images. For our study, we use the profile-fitted photometry measurements of the W1 (3.4$\mu$m), W2 (4.6$\mu$m) and W3 (12$\mu$m) bands, flagged {\bf as} \textsf{w1mpro}, \textsf{w2mpro} and \textsf{w3mpro} in the All-WISE Source Catalog. The associated errors are flagged as \textsf{w1sigmpro}, \textsf{w2sigmpro} and \textsf{w3sigmpro}.

\subsection{Photometric redshifts catalogs}

In order to have access to redshift information, we use the union of two full-sky photometric redshift catalogs: 2MPZ and WISExSCOS.
\\
\\
The catalog 2MPZ \citep{bilicki14} is a cross-match of sources from the WISE, SuperCOSMOS and 2MASS catalog of extended sources. The 2MPZ catalog contains around $10^6$ nearby galaxies, with spectroscopic redshifts for one third of the galaxies and photometric redshifts for the other two thirds. The median redshift is $z_{\mathrm{med}}\sim0.08$, and the redshift dispersion is $\sigma_z\sim0.012$. We refer the reader to \cite{bilicki14} for more information about the construction of the catalog, including the discussion on the purity obtained after the cuts performed to clean the galaxy catalog from stars, blending or AGN contamination.
\\
\\
The catalog WISExSCOS \citep{bilicki16} is an extension of the 2MPZ catalog. It is a cross-match of sources from WISE and SuperCOSMOS, and is deeper than 2MPZ. The median redshift of the galaxies in this catalog is $z_{\mathrm{med}}\sim0.2$, and the redshift dispersion is $\sigma_z\sim0.033$. Here again, we refer the reader to \cite{bilicki16} for more information about the catalog.

\subsection{SDSS}

The SDSS has imaged one quarter of the sky in five optical bands: u, g, r, i and z. We use here the MPA–JHU catalog, from the Max Planck Institute for Astrophysics and the Johns Hopkins University \citep{mpajhu1,mpajhu2}, where are provided star formation rates (SFR) and stellar masses for 1,843,200 galaxies. These data based on SDSS DR8 release are publicly available\footnote{\url{http://sdss.org}} together with all details about the catalog and the computations and fits of the galaxy physical properties. For our study, we extract from the catalogs the median values of SFR and stellar masses, flagged as \textsf{SFR\_TOT\_P50} and \textsf{LGM\_TOT\_P50} respectively.

\section{Selection of galaxy cluster pairs}\label{selectioncluster}

\begin{figure}[!h]
\centering
\includegraphics[width=0.5\textwidth]{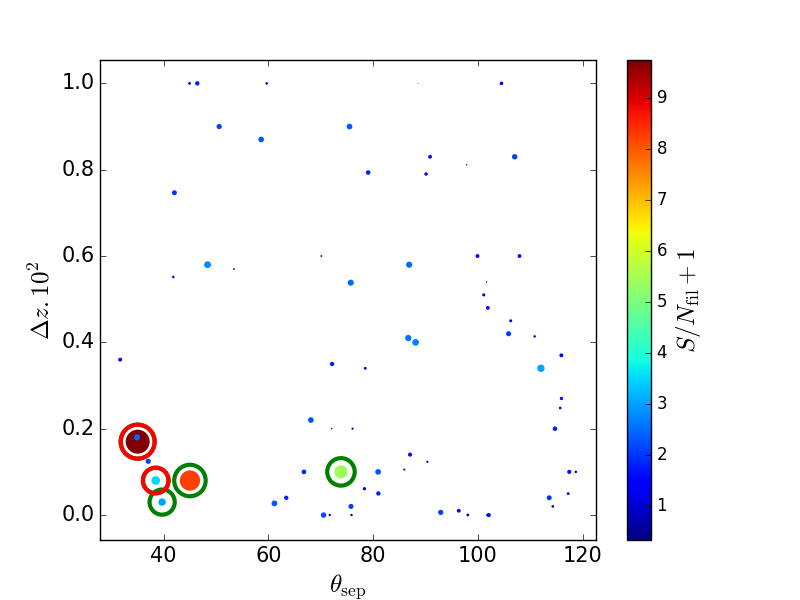}
\caption{\label{space_parameter} Distribution of the 71 cluster pairs in the parameter space $\Delta z$-$\theta_{\mathrm{sep}}$. The areas and colors of the circles depend on $S/N_{\mathrm{fil}}$. The pairs for which $S/N_{\mathrm{fil}}>2$ are surrounded in green. The red surrounded circles indicate the final selection after exclusion of pairs in complex systems.}
\end{figure}

In our analysis, we select the cluster pairs based on the tSZ signal given that it is a priori a more appropriate tracer of the diffuse hot gas. We therefore base our selection of the cluster pairs both on the SZ catalog of clusters and on the signal-to-noise ratio of the tSZ signal between the pairs. In order to define the cluster region, we estimate the cluster extension $r_{500}$\footnote{$r_{500}$ is defined as the radius of the sphere within which the cluster mean mass over-density is 500 times the critical density}, defined as $r_{500}=\left(\frac{1}{500}2G{H_0}^{-2}
E(z)^{-2}M_{500}\right)^{\frac{1}{3}}$ \citep{law}, for the 1681 SZ clusters for which both redshift and mass estimate $M_{500}^{\mathrm{SZ}}$ are available.
\\
\\
Following \cite{planck135}, we apply three conditions to select the galaxy-cluster pairs. First, the two clusters need to be at the same redshift. Second, the distance between the two clusters need to be large enough to avoid blending effects. Finally the significance of the tSZ signal in the inter-cluster
region need to be above $2\sigma$. The two first empirical conditions were proposed by \cite{planck135}: $\Delta z$<0.01, and considering the Planck tSZ map beam of 10 arcmin, 30 arcmin<$\theta_{\mathrm{sep}}$<120 arcmin. Here $\Delta z$ is the redshift difference between the two clusters and $\theta_{\mathrm{sep}}$ is the angular distance separating the two clusters. This corresponds to projected distances between 3 and 40 Mpc. We find that a total of 71 cluster pairs satisfy the two conditions (Fig.~\ref{space_parameter}). This is about three times more pairs than the selection based on clusters from the MCXC catalog (Meta-Catalog of X-ray detected Clusters) \citep{mcxc}, performed in \cite{planck135}. About one third of the clusters in these pairs are Abell clusters \citep{abell}, one third are Planck newly detected clusters, and the others are X-ray clusters from the MCXC catalog or SZ clusters detected by SPT. 
\\
\\
Since our aim is to study and characterize the filamentary gas between pairs of clusters, we consider the tSZ signal significance in terms of the signal-to-noise ratio of the tSZ emission in the filament region, $S/N_{\mathrm{fil}}$. For each cluster pair, we extract patches of the Planck tSZ map of $s=90$ pixels aside, centered on the mean of the two cluster positions, with a patch resolution $\theta_{\mathrm{p}}$ depending on the angular distance $\theta_{\mathrm{sep}}$ separating the two clusters: $\theta_{\mathrm{p}}=(5\times \theta_{\mathrm{sep}})/s$. We rotate the image for convenience, and we mask the tSZ signal at the positions of the Planck Catalog of Compact Sources \citep{pccs} and at ones of the other tSZ sources in the fields, except in an area defined as the region encompassed within $3\times r_{500}$ from the cluster center. Figure \ref{maps} shows such a tSZ map with the red circles representing the $r_{500}$ radii of the clusters, and the white pixels the masked regions. On these patches, we define the area of the potential inter-cluster filament as the cylinder between the minimum of the two $r_{500}$ in the radial direction and between the two $r_{500}$ in the longitudinal direction. We estimate the signal-to-noise $S/N_{\mathrm{fil}}=(y_{\mathrm{m,fil}}-y_{\mathrm{m,bkgd}})/\sigma_{\mathrm{bkgd}}$, where $y_{\mathrm{m,fil}}$ is the mean tSZ signal in the area defining the filament, and $y_{\mathrm{m,bkgd}}$ and $\sigma_{\mathrm{bkgd}}$ are the tSZ signal and the standard deviation of the background area beyond the $3\times r_{500}$ distance from the cluster centers. 
\\
\\
We show in Fig.~\ref{space_parameter} the 71 cluster pairs that satisfy the two first conditions (redshift and angular separations). The areas and colors of the circles display the estimated $S/N_{\mathrm{fil}}$. The surrounded green circles are the pairs for which the signal-to-noise is greater than two. Among these, we choose to discard the cluster pairs belonging to larger and more complex super-clusters such as Shapley \citep{shapley}, or the pair containing the cluster A3395, a system known to host several groups \citep{a3995,planck135}. We also finally remove the pair SPT J0655-5234–SPT J0659-5300, at redshift $z=0.47$, with $S/N_{\mathrm{fil}}=2.03$. The redshift of this system is high, and thus the two catalogues used to study the galaxies in Sect.~\ref{galaxies}, with $z_{\mathrm{med}}\sim0.08$ and $z_{\mathrm{med}}\sim0.2$, lack statistics. A dedicated study of the removed pairs will be done later. Eventually, we focus on two isolated pairs with a significant tSZ signal in their inter-cluster region (red surrounded circles in Fig.~\ref{space_parameter}): the pair A399-A401 at redshift $z\sim0.073$ with $S/N_{\mathrm{fil}}=8.74$ and the newly associated pair A21-PSZ2 G114.09-34.34 at redshift $z\sim0.094$ with $S/N_{\mathrm{fil}}=2.53$. Their main properties are presented in Tab.~\ref{tab}, and the patches of the Planck SZ map are shown Fig.~\ref{maps}.

\begin{table*}[t]
\centering
\begin{tabular}{|c|c|c|c|c|c|c|c|c|c|c|}
\hline
 Cluster & R.A. (deg) & Dec. (deg) & $S/N_{\mathrm{SZ}}$ & $S/N_{\mathrm{d}}$ & $z$ & $r_{500}$ (Mpc) &  $\theta_{\mathrm{sep}}$ (') &$\theta_{\mathrm{sep}}$ (Mpc) & $S/N_{\mathrm{fil}}$ & $S/N_{\mathrm{fil,d}}$\\ \hline \hline
 A399  &  44.45   &   13.05    & 12.96 & 8.85 &  0.072    &     1.20   &       \multirow{2}{*}{35.05} & \multirow{2}{*}{3.0} & \multirow{2}{*}{8.74} & \multirow{2}{*}{7.89} \\ \cline{1-7} 
                            A401      &  44.73    &   13.57     & 19.66 & 9.74 &   0.074    &      1.30    &    &   &  &         \\ \hline \hline
 A21     & 5.15        &  28.67       & 9.36 & 6.19 &  0.094   &   1.07    &  \multirow{2}{*}{38.5}&\multirow{2}{*}{4.2}&\multirow{2}{*}{2.53}&\multirow{2}{*}{2.08} \\ \cline{1-7} 
                            PSZ2 G114.9            & 5.35 &28.05 & 4.78 & 2.63 & 0.095  &   0.93    &          && &    \\ \hline
\end{tabular}
\caption{\label{tab}Main properties of the selected pairs: cluster names, tSZ positions in R.A. and Dec., signal-to-noise ratios of the clusters in the SZ map (SZ cluster database) and in galaxy over-density (Sect.~\ref{overdens}), redshifts, estimated radii $r_{500}$. The last four columns indicate the angular separation in arcmin and in Mpc, and the signal-to-noise ratios of the filaments in the tSZ map (Sect.~\ref{selectioncluster}) and in galaxy over-density (Sect.~\ref{overdens}).}
\end{table*}

\section{Gas properties from tSZ analysis}\label{gas}

\begin{figure}[!h]
\centering
\includegraphics[width=0.5\textwidth]{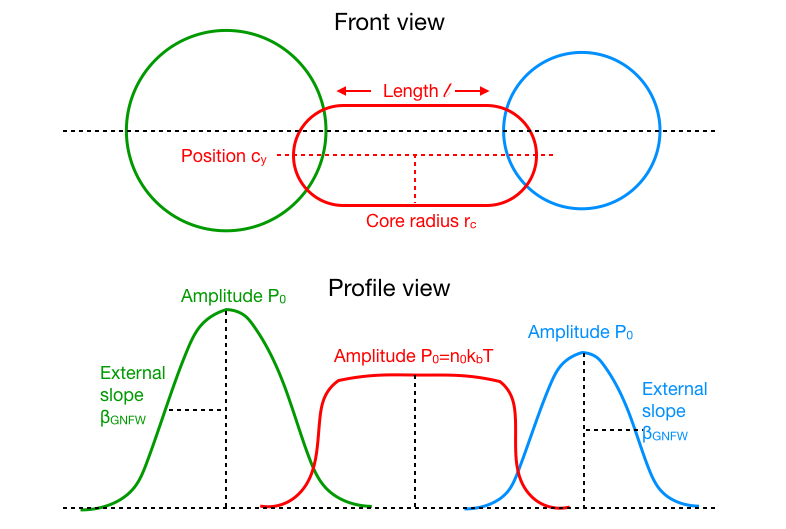}
\caption{\label{schema}Front and profile schematic views of the model: The two clusters in green and blue with two free parameters each, and the inter-cluster filament in red with three free parameters. The length of the filament $l$ is fixed to the distance between the two $r_{500}/2$ of each clusters. A planar background with three free parameters is considered.}
\end{figure}

\subsection{Model}\label{model}
In order to derive the gas properties of the inter-cluster filament, we model the entire system with four components: two clusters, one filament, and a planar background (Fig.~\ref{schema}). To model the galaxy clusters (blue and green components in Fig.~\ref{schema}), we choose to use the spherically symmetric Generalized Navarro Frenk \& White (GNFW) pressure profile \citep{nagai,arnaud,planckgnfw} given by:

\begin{equation}
P\left(\textbf{r}\right)=\frac{P_0}{\left(\frac{\textbf{r}}{r_{\mathrm{s}}}\right)^\gamma\left[1+\left(\frac{\textbf{r}}{r_{\mathrm{s}}}\right)^\alpha\right]^{\left(\beta_{\mathrm{GNFW}}-\gamma\right)/\alpha}},
\end{equation}
\
where $\textbf{r}$ is the radius, $P_0$ is the central pressure, $r_{\mathrm{s}}=r_{500}/c_{500}$ is the characteristic radius, and $\gamma$, $\alpha$, $\beta_{\mathrm{GNFW}}$ respectively the internal ($\textbf{r}<r_{\mathrm{s}}$), intermediate ($\textbf{r}\sim r_{\mathrm{s}}$) and external ($\textbf{r}>r_{\mathrm{s}}$) slopes. All the parameters, but two, are fixed to the ones of the \textit{universal profile} fitted by \cite{planckgnfw} on stacked tSZ clusters. The two parameters we let free are $\beta_{\mathrm{GNFW}}$ (related to the cluster extension) and $P_0$ (related to the Compton parameter $y$ amplitude). These two parameters can be degenerate with each other, and with the parameters of the filament such as its length or its pressure (see Sect.~\ref{resultpair1}).
\\
\\
The GNFW pressure profile was specifically developed to model the pressure distribution in galaxy clusters. As such it cannot be applied to filaments. In the absence of any physically motivated or established model of the inter-cluster filamentary gas, we choose a simple isothermal $\beta$-model \citep{beta}, with a cylindrical geometry, to describe the pressure distribution in the radial direction of the filament:

\begin{equation}
n_{\mathrm{e}}\left(\textbf{r}\right)=\frac{n_{\mathrm{e},0}}{\left(1+\left(\frac{\textbf{r}}{r_{\mathrm{c}}}\right)^2\right)^{\frac{3}{2}\beta}},
\end{equation}
\
where $\textbf{r}$ is the radius, $n_{\mathrm{e},0}$ is the central electron density in the filament, $r_{\mathrm{c}}$ the core radius and $\beta$ the slope, fixed to 4/3 to model a non-magnetised filament \citep{ostriker}. Three parameters are let free: the central pressure in the filament $P_0=n_{0}\times k_{\mathrm{b}}T$ (where $T$ is the electron gas temperature), the core radius $r_{\mathrm{c}}$ (filament extension in the radial direction), and the position of the filament in the radial direction $c_{\mathrm{y}}$ (expressed in percentage of the map size). The length of the filament in the horizontal direction $l$ is fixed here to the distance between the two $r_{500}/2$ of each clusters to avoid any degeneracies between the parameters of the filament and those of the two clusters, and thus avoid any bias in the measurements of the clusters' pressures.
\\
\\
Finally, a plane ($f(x,y)=ax+by+c$) is chosen to model the background. It corrects from possible residual gradient emission induced by large-scale contamination to the tSZ signal such as galactic dust emission.

\subsection{Results for the pair A399-A401}\label{resultpair1}
\begin{figure}[!h]
\centering
\includegraphics[width=0.5\textwidth]{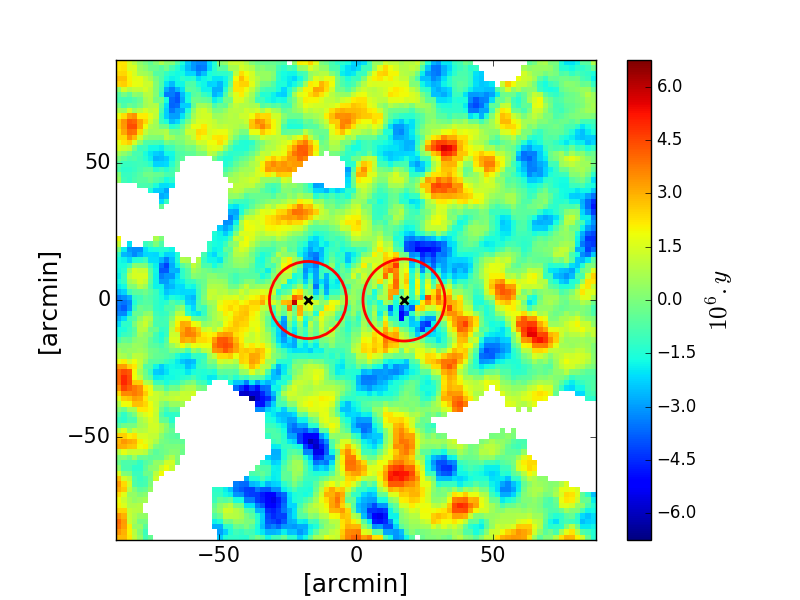}
\caption{\label{res}Residual Planck tSZ map of the pair A399-A401 after subtracting the model (clusters + inter-cluster filaments) with the best-fit parameters from Tab.~\ref{tab2}. The red circles represent the $r_{500}$ radii of each cluster, and the black crosses their central positions.}
\end{figure}

\begin{figure}[!h]
\centering
\includegraphics[width=0.5\textwidth]{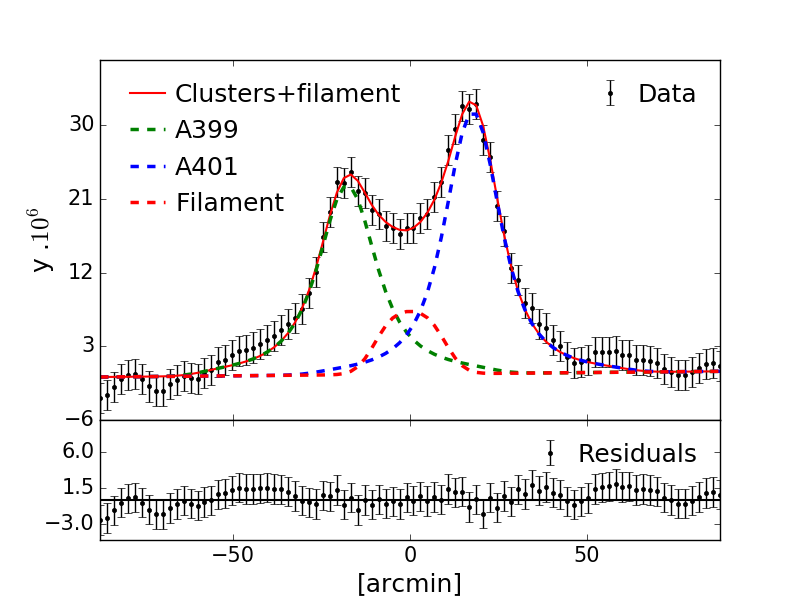}
\caption{\label{resum}Longitudinal cut across A399-A401 system. Top panel: The fuchsia line shows the Planck tSZ data. The red line is the model (clusters + inter-cluster filament) with the best-fit parameters from Tab.~\ref{tab2}. The dotted green, blue, and red lines shows the contributions of A399, A401, and of the inter-cluster filament respectively.Bottom panel: Residuals after subtracting the model (clusters + inter-cluster filament).}
\end{figure}

\begin{figure*}[t]
\centering
\includegraphics[width=\textwidth]{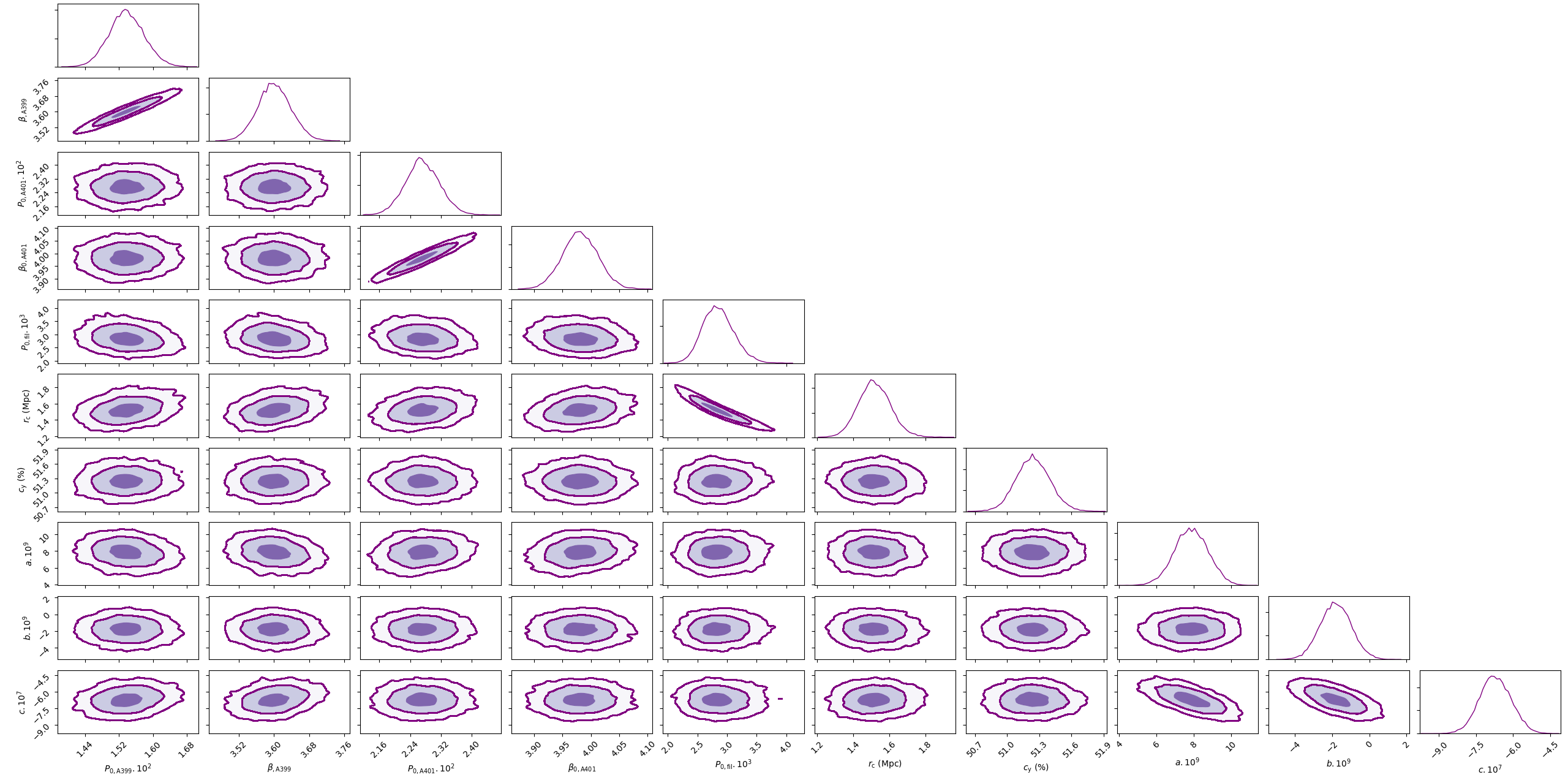}
\caption{\label{triang1} Posterior parameter distributions from the MCMC analysis of the system A399-A401. The diagonal plots show the 1D likelihood of the 11 parameters.
%All the parameters are well constrained, except the length $l$ of the filament in the longitudinal direction. The parameter has reached his maximum value and is evolving with the same likelihood. This is due to the degeneracies between $l$ and the free parameters of the two clusters. This parameter can not be extend as we want to ensure a good quality measurement of the parameters of the two clusters.
}
\end{figure*}

We perform a Monte Carlo Markov Chain (MCMC) analysis using the \texttt{python} algorithm \texttt{emcee} \citep{emcee} on the pair A399-A401 to fit the 
10 free parameters to the most recent Planck tSZ map \citep{szmap}. The resulting posterior parameter distributions are shown in Fig.~\ref{triang1}. 
\\
\\
All the parameters are well constrained. We first note the total absence of degeneracies between the three parameters of the background and the physical parameters of the model. This shows that for the cluster pair A399-A401 the physical properties that we measure, such as the pressures in the clusters or in the filament, are not biased by a potential large-scale contamination.
An interesting point when looking at the correlations between the 10 parameters is the degeneracies between the two parameters of the cluster model: $\beta_{\mathrm{GNFW}}$ and $P_0$. This reflects the fact that a high tSZ amplitudes can be obtained by the combination of a high external slope and a small cluster extension. We do not see any obvious degeneracies with other parameters. Similarly to the cluster we also note a degeneracy between the two physical parameters of the filament central pressure and radial extension: $P_0=n_{0}\times k_{\mathrm{b}}T$ and $r_{\mathrm{c}}$.
\\
\\
We find the best-fit median central pressure in the filament, from the MCMC analysis, is $P_{0}=(2.84\pm0.27)\times10^{-3} \mathrm{keV.cm}^{-3}$, and the best-fit radius is $r_{\mathrm{c}}=1.52\pm0.09$ Mpc that is $r_{\mathrm{c}}=17.6\pm1.1$ arcmin. The filament's pressure obtained from our tSZ-only analysis is in perfect agreements with the value obtained with the density and the temperature measurements from the tSZ/X-rays analysis by \cite{planck135}: $P_{0}=(2.6\pm0.5)\times10^{-3} \mathrm{keV.cm}^{-3}$. Assuming the most accurate temperature measured by \cite{fujita} with Suzaku in X-rays, $k_{\mathrm{b}}T_{\mathrm{X}}=6.5\pm0.5 \mathrm{keV}$, we estimate the central density in the filament to $n_{\mathrm{0}}=(4.3\pm0.7)\times10^{-4} \mathrm{cm}^{-3}$, which is here again in full agreement with the value $n_{\mathrm{0}}=(3.7\pm0.2)\times10^{-4} \mathrm{cm}^{-3}$ derived by \cite{planck135} from their combined analysis of tSZ and X-ray signals.
We compute the model (clusters + filament) using the best-fit parameters derived from the MCMC (see Tab.~\ref{tab2}). The reduced chi-square value obtained from the comparison of the model to the tSZ map is $\chi^2=0.97$. We show the residual tSZ signal after subtracting the model (clusters + filament) in Fig.~{\ref{res}}. It exhibits no significant tSZ emission between the clusters. Finally, we illustrate the good agreement between the model and the data by showing in Fig.~\ref{resum} the model superimposed with the longitudinal cut in the tSZ map, across the two clusters.

\begin{table*}[t]
\centering
\begin{tabular}{|c|c|c|c|c|c|c|}
\hline
 \multicolumn{2}{|c|}{A399} & \multicolumn{2}{c|}{A401} & 
 \multicolumn{3}{c|}{Filament} \\ \hline \hline
 
 $P_{0} (\mathrm{keV.cm}^{-3})$ & $\beta_{\mathrm{GNFW}}$ &  $P_{0} (\mathrm{keV.cm}^{-3})$ & $\beta_{\mathrm{GNFW}}$ &
 $P_0 (\mathrm{keV.cm}^{-3})$ & $r_{\mathrm{c}}$ (Mpc) & $c_{\mathrm{y}}$ (\%) \\ \hline
 $(1.54\pm0.04)\times10^{-2}$   &   $3.60\pm0.04$    & $(2.27\pm0.04)\times10^{-2}$   &   $3.98\pm0.03$     & $(2.84\pm0.27)\times10^{-3}$ & $1.52\pm0.09$ & $51.24\pm0.15$ \\ \hline
\end{tabular}
\caption{\label{tab2}Best-fit parameters of the model derived from the MCMC. The best values are the median of the parameters distributions, and the error-bars are computed with the $16^{\mathrm{th}}$ and the $84^{\mathrm{th}}$ percentiles.}
\end{table*}

\subsection{Results for the pair A21-PSZ2 G114.90-34.35}
\begin{figure}[!h]
\centering
\includegraphics[width=0.5\textwidth]{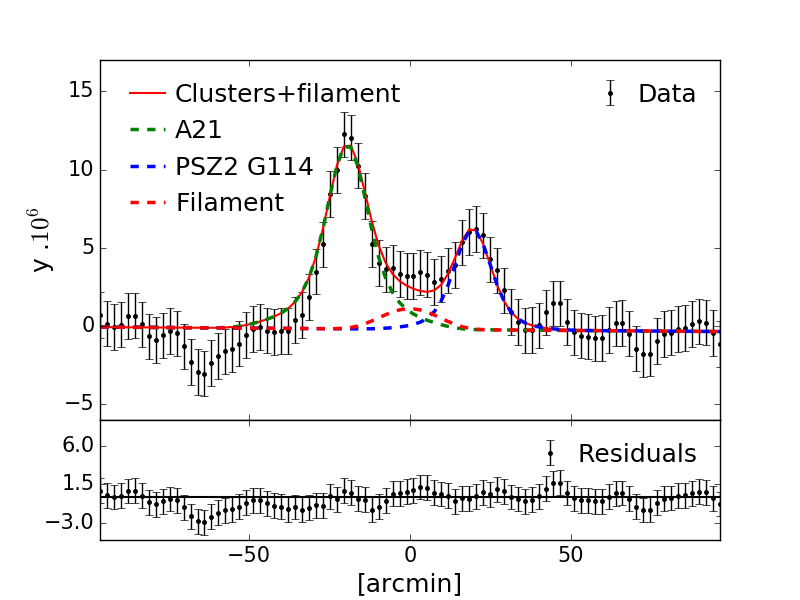}
\caption{\label{resum2} Longitudinal cut across the A21-PSZ2 G114.90-34.35 system. The labels are the same as in Fig.~\ref{resum}.}
\end{figure}

The newly associated pair A21-PSZ2 G114.90-34.35 shows hints of a tSZ signal at $S/N_{\mathrm{fil}}\sim2.5$, associated with the inter-cluster region. We perform the same analysis of the tSZ map as for A399-A401. Similarly, we find that the parameters $P_0$ and $\beta_{{\mathrm GNFW}}$ of the two clusters are degenerate. Considering the lower significance of the tSZ signal in this case, it is difficult to fit the parameters of the model with the MCMC analysis. Therefore, we further fixed the extension radius of the filament $r_{\mathrm{c}}$. As we expect this parameter to be lower or equal to the clusters' extensions, we set it to the smallest $r_{500}$ of the two clusters, here $r_{\mathrm{c}}$=0.92 Mpc or $r_{\mathrm{c}}$=8.5 arcmin. The deduced gas pressure of the filament is $P_{0}=1.6^{+0.7}_{-0.3}\times10^{-3} \mathrm{keV.cm}^{-3}$. In Fig.~\ref{resum2}, we show the contributions of the three components of the system (clusters + inter-cluster filament) computed with the best-fit parameters from the MCMC analysis of the tSZ map and over-plotted on the longitudinal cut across the map. The reduced chi-square is $\chi^2=0.96$ indicating the good agreement between the model and the data. 

\section{Galaxy properties in the cluster pairs}\label{galaxies}

X-ray and tSZ observations have allowed the detection of diffuse gas in between the clusters of the systems A399-A401 and A21-PSZ2 G114.90-34.35, and we have constrained the filament's properties with the tSZ signal from Planck. We study, here, the galaxies in the three components (clusters and filaments) of the two systems, and investigate the possible differences between their properties (types, star formation rates, and stellar masses). We focus on the pair A399-A401 and summarize the main results of the pair A21-PSZ2 G114.90-34.35.

\subsection{Selection for A399-A401}

\begin{figure}[!h]
\centering
\includegraphics[width=0.5\textwidth]{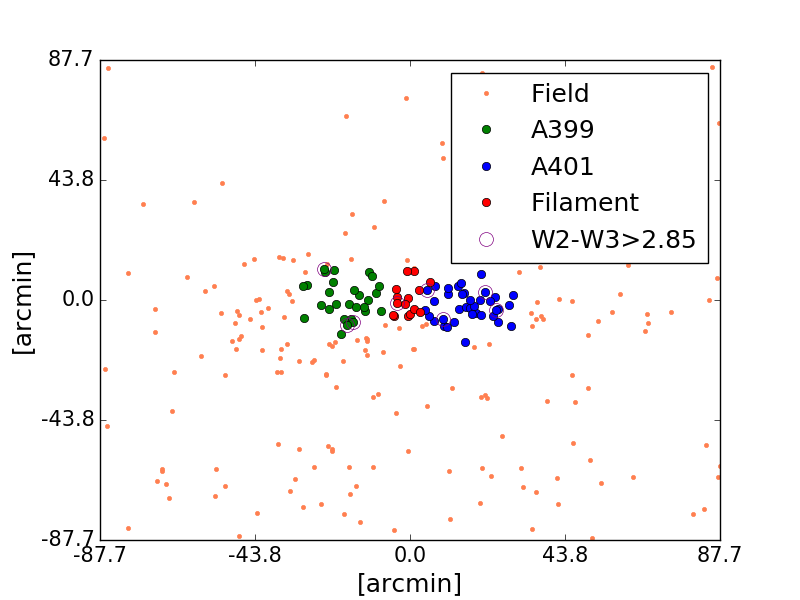}
\caption{\label{selection} The orange dots represent the galaxies in the field of A399-A401 in the redshift range $0.068<z<0.078$. The green dots are the galaxies selected to be in the galaxy cluster A399, the blue ones are the galaxies selected to be in A401, and the red dots are those of the inter-cluster filament. The purple circles indicate galaxies with W2-W3>2.85.}
\end{figure}

We use the photometric bands and photometric redshifts from SuperCosmos, WISE, and 2MASS. We follow \cite{bilicki16} who assume that the K band of 2MASS and the W1 band of WISE are equivalent for low redshift galaxies. In that way, the union catalog provides us with redshifts for all galaxies with W1<17. We cross-match the resulting union catalog and the All-WISE Source Catalog (following \cite{bilicki14}, the matching distance is set to <3 arcsec). We then retrieve the source magnitudes in the bands W1, W2, and W3 and we apply the k-correction factor $-2.5log(1+z)$. 
\\
\\
We first select all galaxies in the redshift range $\sim$0<$z$<$\sim$0.5 within the field of the Planck tSZ map (see Fig.~\ref{maps}). This first selection represents 6487 galaxies. We then refine the selection in order to focus on galaxies likely to belong to the A399-A401 system. To do so, we select galaxies in the range $0.068<z<0.078$ which corresponds to redshifts comprised between the lowest and highest redshifts of the cluster system plus and minus the velocity dispersion of the two clusters \citep{oegerle}. For simplicity and coherence with the tSZ analysis, we assume that member galaxies are defined as those within an area of radius $r_{500}$, centered on the tSZ cluster positions. The galaxies which belong to the inter-cluster filament are defined to be spatially contained between the positions $c_{\mathrm{y}}$ plus or minus the core radius $r_{\mathrm{c}}$, both parameters fitted with the MCMC in Sect.~\ref{gas}. Galaxies outside these regions are considered as field galaxies. We show in Fig.~\ref{selection} the selected galaxies in the different components of the system A399-A401. The orange dots are for the field galaxies whereas the green, blue, and red dots represent galaxies belonging respectively to A399, A401, and to the inter-cluster filament.

\subsection{Galaxy over-densities}\label{overdens}

\begin{figure}[!h]
\centering
\includegraphics[width=0.5\textwidth]{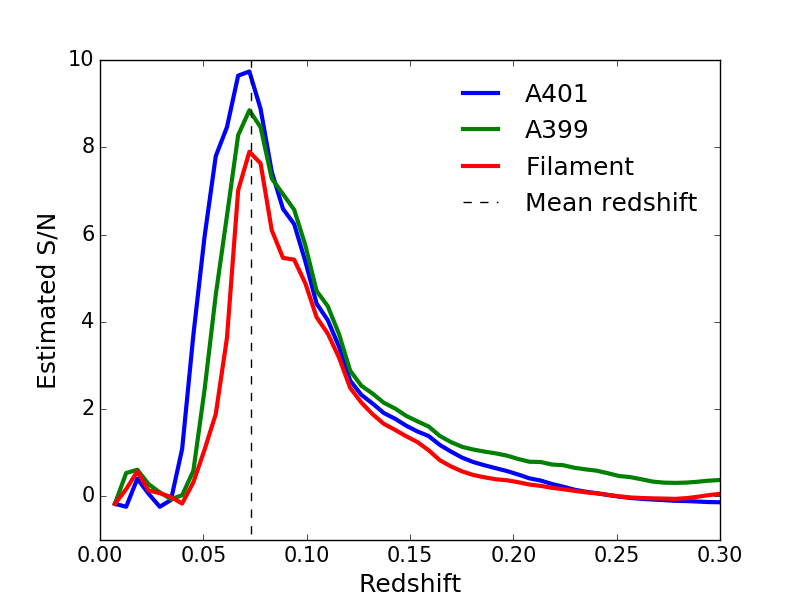}
\caption{\label{snr} Distributions of the signal-to-noise ratios of over-densities for the three components of the system A399-A401. The blue line corresponds to cluster A401, the green line to A399 and the red one to the inter-cluster filament region.}
\end{figure}

We have implemented a three-dimensional Delaunay Tesselation Field Estimator (3D DTFE in \texttt{python}) \citep{dtfe3,dtfe2,dtfe1} to estimate a 3D galaxy density in the field of A399-A401. We compute the over-density based on all galaxies, such that $\sim$0<$z$<$\sim$0.5, within the field of the Planck tSZ map. For each redshift bin of width $\delta z=0.005$, we compute the mean over-density in the three regions defining the components of the system (i.e. clusters and filament). We compare the mean density in the galaxy cluster regions and in the filament to the mean density of the background and its standard deviation. In Fig.~\ref{snr}, we show the distribution of the over-density signal-to-noise as a function of redshift for the three components of the system. It is clear that the signal-to-noise ratios in the regions of the two galaxy clusters and the one of the inter-cluster filament peak at the mean redshift of the system. We conclude that we detect a structure with a confidence, $S/N_{\mathrm{fil}}\sim8$, comparable to the tSZ signal.

\subsection{Optical properties}

\begin{figure}[!h]
\centering
\includegraphics[width=0.5\textwidth]{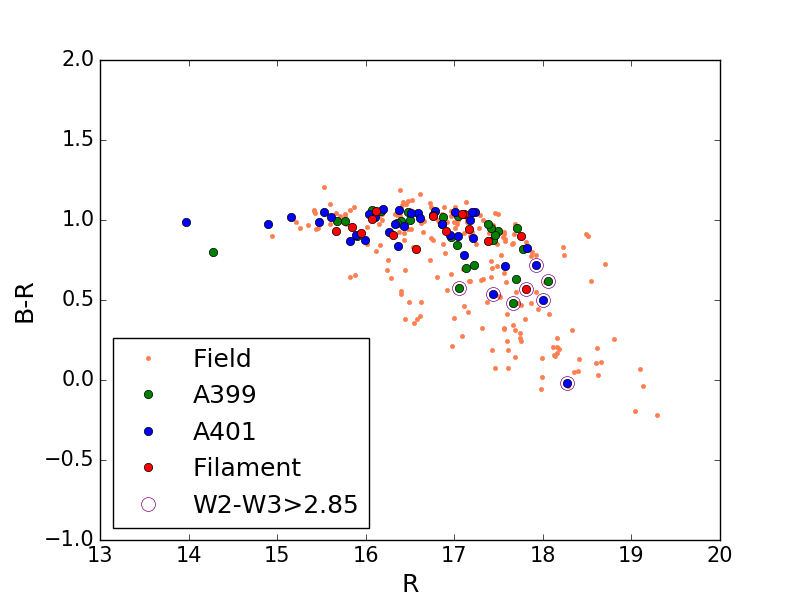}
\caption{\label{CMD} Color-magnitude diagram using B and R photometric bands from SuperCOSMOS. The labels are the same as those in Fig.~\ref{selection}.}
\end{figure}

A standard method to identify collapsed structures (e.g. galaxy clusters) from galaxy optical properties is to search for a red sequence \citep{visv,gladders,rykoff}. Indeed, galaxies in clusters are expected to be red and dead at low redshift, with the 4000\AA-break the main feature in their spectra. The colors of the galaxies in a cluster computed with two magnitudes surrounded this break will be roughly the same, and a linear feature will form in the color-magnitude diagram, called the red sequence. 
\\
\\
As the 4000\AA-break is between the magnitudes B and R for low redshift galaxies, we use the photometric bands of SuperCOSMOS to produce the color-magnitude diagram B-R vs R, for the galaxies in the system A399-A401. The resulting plot is shown Fig.~\ref{CMD}. We note that the galaxies in the three components of the system tend to align on a red sequence. An interesting point is that the galaxies in the filament region are also following the same red sequence as the one in the clusters. We will investigate that point and confirm that trend by estimating galaxy types with infrared properties.

\subsection{Infrared properties}

\begin{figure}[!h]
\centering
\includegraphics[width=0.5\textwidth]{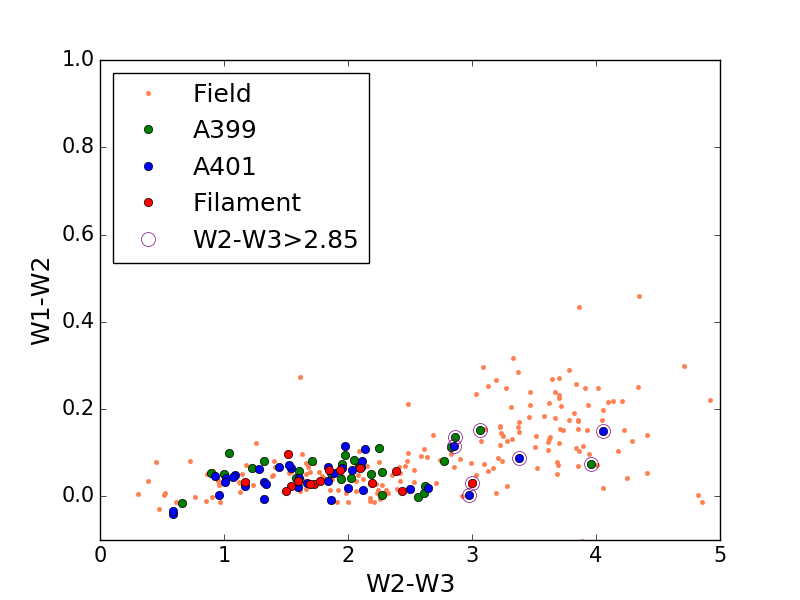}
\caption{\label{CCD} Color-color diagram from the photometric bands W1, W2 and W3. The labels are the same as those in Fig.~\ref{selection}.}
\end{figure}

\begin{figure}[!h]
\centering
\includegraphics[width=0.5\textwidth]{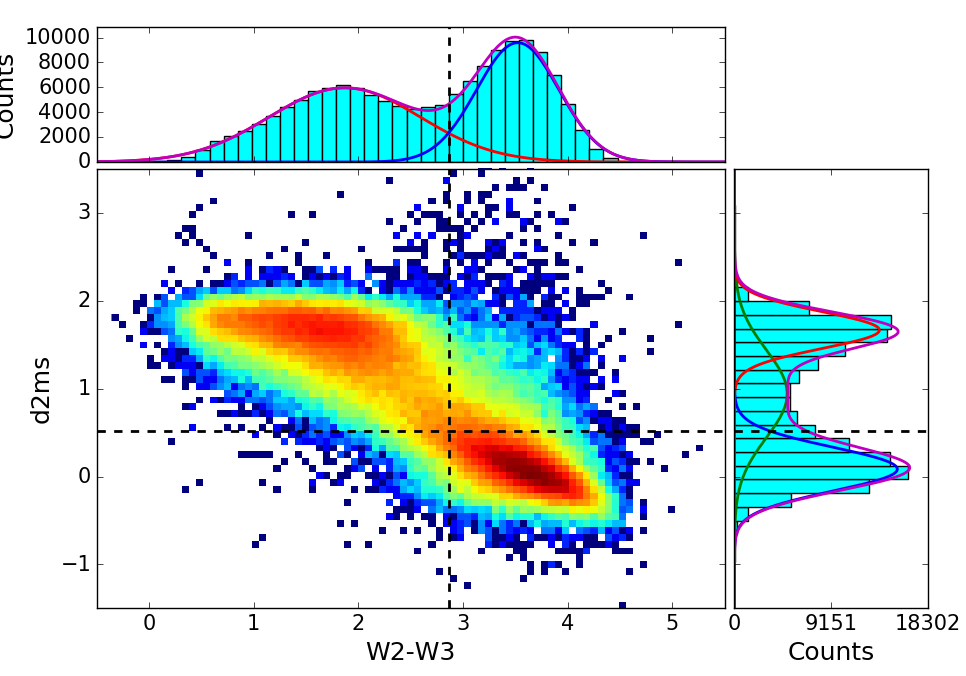}
\caption{\label{d2ms} 2D histogram of the projected distance to the main sequence in the SFR/mass plane vs. the W2-W3 color, for the 148 685 galaxies of the SDSS MPA-JHU DR8 field cross-matched with the union catalog of photometric redshifts. Top: Histogram of the color W2-W3. We have fitted two Gaussians for the early-type galaxies in red, and for the late-type galaxies in blue. Right: Histogram of the projected distance to the main sequence. We have fitted three Gaussians: red and dead (in red), star-forming (in blue), and green valley (in green) galaxies. We set a threshold at W2-W3=2.85 (solid black line) to separate star-forming galaxies.}
\end{figure}

As shown in \cite{lee}, there are two main populations of galaxies having different properties in infrared: early-type and late-type galaxies. The early-type galaxies are massive and passive, composed mostly of elliptical E and spiral S0, which lay on a optical red sequence. They are "red and dead" or in the green valley (transiting from star-forming to passive). Late-type galaxies are mainly star-forming, blue, and spiral galaxies. \cite{wise} have used color-color diagrams based on the W1-W2 and W2-W3 near-infrared colors from WISE to segregate between galaxy morphologies: elliptical or spiral galaxies. Elliptical galaxies were found to be mainly located between 0.5<W2-W3<1.5 and -0.1<W1-W2<0.3, and spirals between 1<W2-W3<4.5 and -0.1<W1-W2<0.7 (see Fig.~12 in \cite{wise}). In a similar way, we use the bands W1, W2, and W3 and we show that all galaxies in the A399-A401 system (clusters + filament) are located in the two regions defining elliptical and spiral galaxies (see Fig.~\ref{CCD}).
\\
\\
We further use the W2-W3 color to segregate between late-type and early-type galaxies. To do so, we cross-match the SDSS MPA-JHU DR8 catalog with the union catalog of photometric redshifts. For each of the 148 685 galaxies, we use the SFR and stellar mass provided in the catalogs to evaluate the projected distance, noted d2ms, in the SFR/mass plane to the SFR/mass relation of the main sequence for star-forming galaxies given by \cite{mpajhu2}. We show in Fig.~\ref{d2ms} the histogram of the distances as compared with the histogram of the W2-W3 colors for the same galaxies. The black dotted line in Fig.~\ref{d2ms} represents the W2-W3=2.85 color-cut above which the galaxies are star-forming ones, with projected distances to the main sequence d2ms $\sim$ 0. In all the plot, we display the star-forming galaxies with purple circles.
The threshold at W2-W3=2.85 between early and late-type galaxies agrees well with the one found by \cite{alatalo} between the green valley (early types) and the star forming galaxies (late types). The galaxies which align on the red sequence in Fig.~\ref{CMD} are the same than the ones belonging to the early-type population (as defined by W2-W3<2.85) in Fig.~\ref{CCD}. \\
\\
These consistent results indicate that there is no obvious segregation between the galaxies in the clusters and in the inter-cluster filament. Most of the galaxies in the three components of the A399-A401 system are passive, mainly composed of elliptical E and spiral S0.

\subsection{SFRs and stellar masses}
\begin{figure}[!h]
\centering
\includegraphics[width=0.5\textwidth]{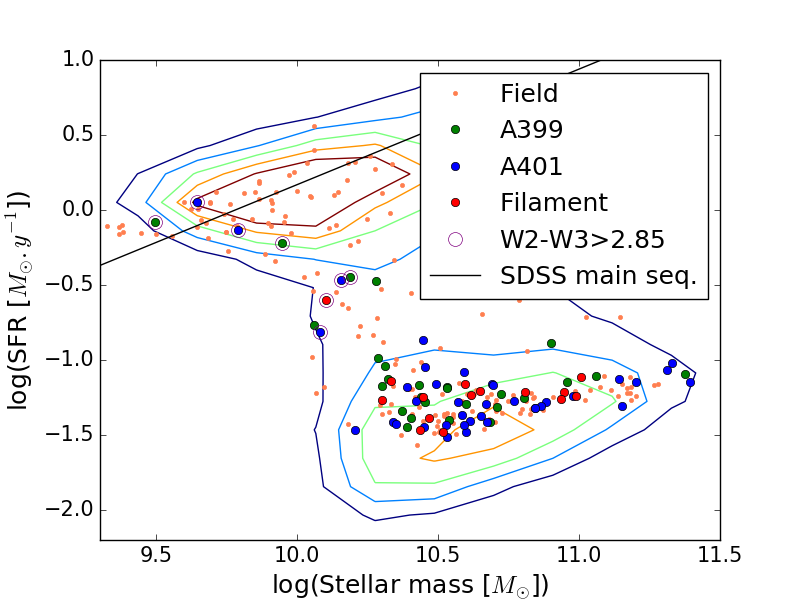}
\caption{\label{SFR-m} SFRs versus stellar masses for the galaxies in the field of A399-A401. The labels are the same as those in Fig.~\ref{selection}. The black line corresponds to the main sequence fitted by \cite{mpajhu2} on the SDSS MPA-JHU galaxies. The contours represent the galaxies from the cross-match between the union catalog and the SDSS MPA-JHU DR8 catalog, in the redshift range $0.068<z<0.078$.}
\end{figure}

The infrared luminosity traces the SFRs and the stellar masses of galaxies \citep{calzetti,cluver}, as the dust heated by young stars (indicators of the SFR) and old stars (indicators of the stellar mass) re-emits in the infrared wavelengths. \citet{cluver} have fitted correlations between the SFR and the luminosity in the W3 band for star-forming galaxies, and between the stellar mass, the luminosity in the W1 band, and the color W1-W2. 
\\
\\
In the following, we estimate the SFRs and stellar masses for passive galaxies in order to confirm the results obtained from the optical and infrared color properties. To do so, we develop a method based on a machine learning algorithm\footnote{\texttt{MLPRegressor} in \texttt{scikit-learn} library in \texttt{python}}, trained with the SDSS MPA-JHU DR8 catalog cross-matched with the union catalog of photometric redshifts (Bonjean et al. in prep.). To estimate the SFRs with the neural network, we give in input the luminosity in the W3 band, the colors W1-W2 and W2-W3, the redshifts, and the SFR from the SDSS MPA-JHU DR8 catalog. For the stellar masses, we give in input the luminosity in the W1 band, the colors W1-W2 and W2-W3, and the stellar masses from the SDSS MPA-JHU DR8 catalog.
%We tested different configurations of layers and we concluded that one hidden layer of 30 neurons were enough. We trained the neural network with 148 685 galaxies from the cross-matched catalog. We tested our method, by dividing the catalog of galaxies in two samples: 140 000 galaxies to train the algorithm and 8 685 galaxies to test it. Dividing in that way the sample, we roughly keep the same order of magnitude of the fraction of trained vs. estimated galaxies. The resulting tests on the machine learning algorithm are shown Fig.~\ref{ML}, where we plot the estimated SFRs and stellar masses vs. the SFRs and the stellar masses available in the SDSS catalog for the galaxies of the test sample. 
\\
\\
We estimate with the neural network, the SFRs and the stellar masses of the galaxies in the system A399-A401, and compare them, in Fig.~\ref{SFR-m}, to the ones of the SDSS field and to the main sequence of star-forming galaxies fitted by \cite{mpajhu2}. The contours represent the galaxies of the SDSS field used to train the machine learning algorithm on the same slice of redshift than the selected field, $0.068<z<0.078$. We confirm here that almost all the galaxies in the three components of the system are passive. We also find a very good agreement between estimated early-type galaxies with the proposed threshold at W2-W3>2.85, and the location of active galaxies in the main sequence diagram, as in the color-color and in the color-magnitude diagrams.

%\begin{figure*}[t]
%\centering
%\includegraphics[width=0.5\textwidth]{ML_SFR}\includegraphics[w%idth=0.5\textwidth]{ML_MASS}
%\caption{\label{ML} Results of the machine learning algorithm on 8 685 galaxies from the SDSS MPA-JHU catalog. The algorithm is trained with the other 140 000 galaxies. Left: 2d histogram of estimated SFRs compare to SFRs given by the SDSS catalog. The input values are the luminosities in the W3 band, the redshifts, and the colors of WISE W1-W2 and W2-W3. Right: 2d histogram of estimated stellar masses compare to stellar masses given by the SDSS catalog. The input values are the luminosities in the W1 band and the colors of WISE W1-W2 and W2-W3. The black lines represent the expected 1-1 correlations.}
%\end{figure*}

\subsection{Results for the pair A21-PSZ2 G114.90-34.35}

\begin{figure}[!h]
\centering
\includegraphics[width=0.5\textwidth]{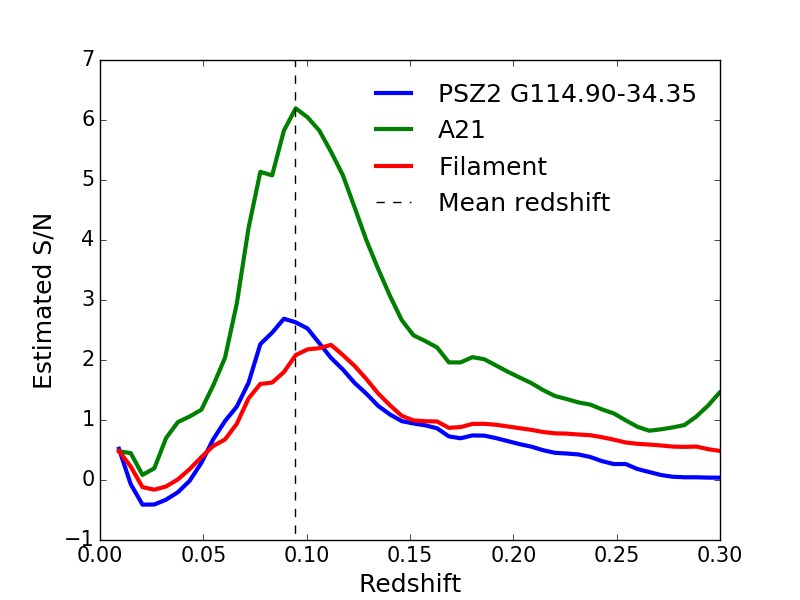}
\caption{\label{snr_a21} Distribution of the signal-to-noise ratios of over-densities for the three components of the system A21-PSZG114.90-34.35. The blue line corresponds to PSZG114.90-34.35, the green line to A21 and the red line to the region between the two clusters.}
\end{figure}

We compute the 3D density field in the region of the pair A21-PSZ2 G114.90-34.35. The significance of an over-density of galaxies between the two clusters is lower than in the A399-A401 case (see Fig.~\ref{snr_a21}). We find, in the inter-cluster region, an over-density with a signal-to-noise ratio of about 2.5. However due to the lack statistics for this cluster system, we cannot study the galaxy properties and conclude on their nature. 

\section{Discussion}\label{discussion}

First observations of the pair A399-A401 in the X-rays with the Einstein telescope \citep{ulmer} did not show any evidence of signal between the two clusters \citep{kara,ulmer} because of the high background noise. It is only with ASCA and ROSAT that \cite{fujita3} and \cite{fabian} detected an excess of X-ray emission that they associated with a diffuse gas in between the two clusters. Given that the two clusters, A399 and A401, lack cooling-flows and contained diffuse radio halos \citep{murgia}, the gas between the two clusters was interpreted as the result of a post-merger \citep{fabian}. Alternatively, \cite{fujita3} suggested a pre-merger scenario to explain the origin of the X-ray emission, since the temperatures of the gas and the two clusters were of the same order. Another explanation to the presence of a filament between the two clusters is that the whole system was formed at the same time and that the gas in the inter-cluster region is a relic of a cosmic filament, as proposed by \cite{fujita2}. The higher quality XMM-Newton data from the four pointing observations of A399-A401 permitted to measure the gas properties of the two clusters and of the filament between them \citep{sakelliou}. This revealed both the absence of big anomalies in the temperature profiles and the relaxed nature of the clusters, with no significant offset between the positions of their BCGs and the central positions of the X-ray emissions.
\cite{sakelliou} further interpreted the radio halos in the two clusters as relics of past minor mergers in each cluster. All these evidences, strongly favoured a pre-merger scenario that was reinforced by the Suzaku observations conducted by \cite{fujita} who found a high metallicity in the inter-cluster region, 0.2 $Z_{\odot}$, of the same order of the metallicity in the clusters, indicative of superwinds that transported metals from the clusters to the inter-cluster region. 
\\
\\
In order to investigate the different scenarii and discriminate between them, it is necessary to fully characterise the properties of the inter-cluster filament. In the present study, we thus complete the analysis of the system A399-A401 by exploring the tSZ signal and the galaxy properties. We have confirmed the presence of gas between the two clusters A399 and A401 with a significance of order 8.5$\sigma$ from the Planck tSZ data alone. We have measured the pressure of the inter-cluster gas, $P_{0}=(2.8\pm0.27)\times10^{-3} \mathrm{keV.cm}^{-3}$, in agreement with the pressure computed with the density and the temperature fitted from the combines tSZ/X-rays analysis by \cite{planck135}, $P_{0}=(2.6\pm0.5)\times10^{-3} \mathrm{keV.cm}^{-3}$. Using the most accurate gas temperature of \cite{fujita}, $k_{\mathrm{b}}T_{\mathrm{X}}=6.5\pm0.5 \mathrm{keV}$, we have further estimated a central density of $n_{\mathrm{0}}=(4.3\pm0.7)\times10^{-4} \mathrm{cm}^{-3}$, in agreement with \cite{planck135}.
\\
\\
Note that \cite{fujita2} have also studied the geometry of the filament between A399 and A401. They consider a uniform density along the line of sight and estimate a Compton parameter in the filament, $y=14.5\pm1.8{\left(\frac{l}{\mathrm{Mpc}}\right)}^{0.5}\times10^{-6}$, where $l$ is an effective depth of the filament along the line of sight. Comparing the obtained Compton parameter with a weight-averaged $y$ parameter in the tSZ map (roughly estimated at $y=14-17\times10^{-6}$ in \cite{planck135}) they deduce an effective density $n_0=3.1\times10^{-4} \mathrm{cm}^{-3}$ and the effective depth $l=1.1$Mpc. Akamatsu et al. compare this depth to the size of the filament in the radial direction, $\sim2.6$Mpc (compatible with our result, a size of $3.0\pm0.2$Mpc), and conclude that the filament is flattened. Following their method, we focus on the very central region of the filament (within 2' of the longitudinal axis) and estimate a Compton parameter $y=17.2\pm1.3{\left(\frac{l}{\mathrm{Mpc}}\right)}^{0.5}\times10^{-6}$, with $k_{\mathrm{b}}T_{\mathrm{X}}=6.5\pm0.5$keV and $n_0=3.3{\left(\frac{l}{\mathrm{Mpc}}\right)}^{-0.5}\times10^{-4}\mathrm{cm}^{-3}$ (see their Fig.~7). We compare this to the value of the tSZ map, $y=22.2\pm1.8\times10^{-6}$, and estimate an effective depth $l=1.7\pm0.5$Mpc. This value suggests that the shape of the filament is consistent with a cylinder, compatible with our hypothesis. However, we note that these computations of the filament depth $l$ strongly depends on the model of the electron density and on the assumed value of $y$.
\\
\\
The re-analysis of the Suzaku data by \cite{fujita2} has shown hints of a shock in the direction parallel to the one linking the two clusters. Such a shock would be incompatible with a merger scenario of two clusters only, since numerical simulations predict shocks in the radial direction \citep{akahori}. This suggests the pre-existing cosmic filament, a hypothesis supported by \cite{planck135}. Our study is complementary and brings additional information on the scenario. The tSZ emission traces and detects diffuse gas, with a density one order of magnitude below the mean densities in the clusters. In addition, we have detected a galaxy over-density between the clusters A399 and A401 with a significance of order 8$\sigma$. The colour-colour and colour-magnitude analyses of the galaxies selected in the system show that they are passive, early type, and red and dead. We do not observe any segregation between the galaxies belonging to the three components of the system. The properties of the galaxies in the A399-A401 system are those of the typical populations of galaxies in clusters or in dense collapsed structures. This contrasts with the results showing a large fraction star-forming galaxies in intermediate-density environments such as filaments \citep{gallazzi,louise}. It suggests that the mechanisms by which galaxies can undergo an enhancement of star formation (e.g. mergers, harassment, ram pressure, etc.) are less/not efficient in the A399-A401 system.

\section{Conclusion}\label{conclusion}

We have performed a tSZ based selection of galaxy-cluster pairs showing hints of tSZ signal potentially associated with inter-cluster filaments. Among the 71 pairs satisfying our selection criteria on redshift and angular distance separations, we have selected the systems at the highest significance: A399-A401 at redshift $z=0.073$ with $S/N_{\mathrm{fil}}=8.74$ and A21-PSZ2 G114.90-34.35 at redshift $z=0.094$ with $S/N_{\mathrm{fil}}=2.53$. For these two systems, we have performed a multi-wavelength analysis that allowed us to constrain the gas properties, to obtain the galaxy types, and to measure their SFRs and stellar masses, in the three components of the system: the two clusters and the inter-cluster filament. 
\\
\\
For the most significant pair A399-A401, we measure a gas pressure in the inter-cluster filament, $P_{0}=(2.84\pm0.27)\times10^{-3} \mathrm{keV.cm}^{-3}$, in agreement with previous studies but better constrained. We cannot identify any significant difference in the star formation activity between the galaxies of the clusters and the galaxies of the inter-cluster filament: they all seem passive and red and dead. Our analysis of the galaxy properties weakens the post-merger hypothesis which was already disfavoured by the lack of big offset between the positions of the BCGs and the centers of the X-ray emissions. Our findings rather favour the scenario in which the gas between the two clusters is associated with a cosmic filament. The gas is collapsing, smoothly compressed, and heated by the future collision of the two clusters A399 and A401. The current data are however not deep enough to accurately measure the effect of environmental quenching in the filament that connects the two clusters.
\\
\\
On the newly associated pair of clusters A21-PSZ2 G114.90-34.35, we detect an inter-cluster filament at $2.5\sigma$ from the analysis of the Planck tSZ map and we find the same significance in the 3D galaxy density field. However, we lack statistics to conduct a study on galaxy properties in the three components of this pair of clusters.
\\
\\
Dedicated observations in the X-rays, with higher resolution tSZ instrument, and optical or near-infrared wide field spectroscopic data will be necessary in order to either confirm the presence of inter-cluster filaments from our selection of galaxy-cluster pairs or/and perform a comprehensive combined analysis of their properties to study their origin and link to the cosmic web. 

\begin{acknowledgements}
The authors acknowledge useful comments from an anonymous referee and fruitful discussions with M. Langer, G. Hurier, E. Pointecouteau, and C. de Boni. This work has received support from PI Invitation Program of Institute for Advanced Research, Nagoya University.
This publication is based on observations obtained with Planck (\url{http://www.esa.int/Planck}), an ESA science mission with instruments and contributions directly funded by ESA Member States, NASA, and Canada. It made use of the SZ-Cluster Database operated by the Integrated Data and Operation Center (IDOC) at the Institut d'Astrophysique Spatiale (IAS) under contract with CNES and CNRS. It also used data obtained from the SuperCOSMOS Science Archive, prepared and hosted by the Wide Field Astronomy Unit, Institute for Astronomy, University of Edinburgh, which is funded by the UK Science and Technology Facilities Council. 
This publication makes use of data products from the Wide-field Infrared Survey Explorer, which is a joint project of the University of California, Los Angeles, and the Jet Propulsion Laboratory/California Institute of Technology, funded by the National Aeronautics and Space Administration. This research made use of Astropy, the community-developed core Phyton package \citep{astropy} and matplotlib library \citep{matplotlib}. This project has received funding from the European Research Council (ERC) under the European Union's Horizon 2020 research and innovation programme grant agreement ERC-2015-AdG 695561.
\end{acknowledgements}


\begin{thebibliography}{}

		\bibitem[Abell et al.(1989)]{abell} Abell, G.~O., Corwin, H.~G., Jr., \& Olowin, R.~P.\ 1989, \apjs, 70, 1 
		\bibitem[Aghanim et al.(2009)]{law} Aghanim, N., da Silva, A.~C., \& Nunes, N.~J.\ 2009, \aap, 496, 637 
		\bibitem[Akahori \& Yoshikawa(2008)]{akahori} Akahori, T., \& Yoshikawa, K.\ 2008, \pasj, 60, L19 
		\bibitem[Akamatsu et al.(2017)]{fujita2} Akamatsu, H., Fujita, Y., Akahori, T., et al.\ 2017, \aap, 606, A1
		\bibitem[Alatalo et al.(2014)]{alatalo} Alatalo, K., Cales, S.~L., Appleton, P.~N., et al.\ 2014, \apjl, 794, L13 
		\bibitem[Alpaslan et al.(2014)]{alpaslan} Alpaslan, M., Robotham, A.~S.~G., Driver, S., et al.\ 2014, \mnras, 438, 177 
		\bibitem[Arnaud et al.(2010)]{arnaud} Arnaud, M., Pratt, G.~W., Piffaretti, R., et al.\ 2010, \aap, 517, A92
		\bibitem[Astropy Collaboration et al.(2013)]{astropy} Astropy Collaboration, Robitaille, T.~P., Tollerud, E.~J., et al.\ 2013, \aap, 558, A33 
		\bibitem[Bernardeau \& van de Weygaert(1996)]{dtfe3} Bernardeau, F., \& van de Weygaert, R.\ 1996, \mnras, 279, 693 
		\bibitem[Bersanelli et al.(2010)]{bersanelli} Bersanelli, M., Mandolesi, N., Butler, R.~C., et al.\ 2010, \aap, 520, A4 
		\bibitem[Bilicki et al.(2014)]{bilicki14} Bilicki, M., Jarrett, T.~H., Peacock, J.~A., Cluver, M.~E., \& Steward, L.\ 2014, \apjs, 210, 9 
		\bibitem[Bilicki et al.(2016)]{bilicki16} Bilicki, M., Peacock, J.~A., Jarrett, T.~H., et al.\ 2016, \apjs, 225, 5 
		\bibitem[Brinchmann et al.(2004)]{mpajhu2} Brinchmann, J., Charlot, S., White, S.~D.~M., et al.\ 2004, \mnras, 351, 1151 
		\bibitem[Calzetti et al.(2007)]{calzetti} Calzetti, D., Kennicutt, R.~C., Engelbracht, C.~W., et al.\ 2007, \apj, 666, 870 
		\bibitem[Cavaliere \& Fusco-Femiano(1978)]{beta} Cavaliere, A., \& Fusco-Femiano, R.\ 1978, \aap, 70, 677
		\bibitem[Cluver et al.(2014)]{cluver} Cluver, M.~E., Jarrett, T.~H., Hopkins, A.~M., et al.\ 2014, \apj, 782, 90
		\bibitem[Dolag et al.(2006)]{dolag} Dolag, K., Meneghetti, M., Moscardini, L., Rasia, E., \& Bonaldi, A.\ 2006, \mnras, 370, 656 
		\bibitem[Durret et al.(2016)]{durret} Durret, F., M{\'a}rquez, I., Acebr{\'o}n, A., et al.\ 2016, \aap, 588, A69 
		\bibitem[Eckert et al.(2015)]{eckert} Eckert, D., Jauzac, M., Shan, H., et al.\ 2015, \nat, 528, 105 
		\bibitem[Edwards et al.(2010)]{louise} Edwards, L.~O.~V., Fadda, D., Biviano, A., \& Marleau, F.~R.\ 2010, \aj, 139, 434 
		\bibitem[Elbaz et al.(2007)]{elbaz} Elbaz, D., Daddi, E., Le Borgne, D., et al.\ 2007, \aap, 468, 33 
		\bibitem[Epps \& Hudson(2017)]{epps} Epps, S.~D., \& Hudson, M.~J.\ 2017, \mnras, 468, 2605
		\bibitem[Fabian et al.(1997)]{fabian} Fabian, A.~C., Peres, C.~B., \& White, D.~A.\ 1997, \mnras, 285, L35 
		\bibitem[Fadda et al.(2008)]{fadda} Fadda, D., Biviano, A., Marleau, F.~R., Storrie-Lombardi, L.~J., \& Durret, F.\ 2008, \apjl, 672, L9 
		\bibitem[Foreman-Mackey et al.(2013)]{emcee} Foreman-Mackey, D., Hogg, D.~W., Lang, D., \& Goodman, J.\ 2013, \pasp, 125, 306 
		\bibitem[Fujita et al.(1996)]{fujita3} Fujita, Y., Koyama, K., Tsuru, T., \& Matsumoto, H.\ 1996, \pasj, 48, 191 
		\bibitem[Fujita et al.(2008)]{fujita} Fujita, Y., Tawa, N., Hayashida, K., et al.\ 2008, \pasj, 60, S343 
		\bibitem[Gallazzi et al.(2009)]{gallazzi} Gallazzi, A., Bell, E.~F., Wolf, C., et al.\ 2009, \apj, 690, 1883 
		\bibitem[Gladders \& Yee(2000)]{gladders} Gladders, M.~D., \& Yee, H.~K.~C.\ 2000, \aj, 120, 2148 
		\bibitem[G{\'o}rski et al.(2005)]{healpix} G{\'o}rski, K.~M., Hivon, E., Banday, A.~J., et al.\ 2005, \apj, 622, 759 
		\bibitem[Guennou et al.(2010)]{guennou} Guennou, L., Adami, C., Ulmer, M.~P., et al.\ 2010, \aap, 523, A21 
		\bibitem[Hunter et al.(2007)]{matplotlib} Hunter, J.~D.\ 2007, Computing In Science \& Engineering, 9, 3
		\bibitem[Hurier et al.(2013)]{guillaume} Hurier, G., Mac{\'{\i}}as-P{\'e}rez, J.~F., \& Hildebrandt, S.\ 2013, \aap, 558, A118
		\bibitem[Karachentsev \& Kopylov(1980)]{kara} Karachentsev, I.~D., \& Kopylov, A.~I.\ 1980, \mnras, 192, 109 
		\bibitem[Kauffmann et al.(2003)]{mpajhu1} Kauffmann, G., Heckman, T.~M., White, S.~D.~M., et al.\ 2003, \mnras, 341, 33
		\bibitem[Lakhchaura et al.(2011)]{a3995} Lakhchaura, K., Singh, K.~P., Saikia, D.~J., \& Hunstead, R.~W.\ 2011, \apj, 743, 78  
		\bibitem[Lamarre et al.(2010)]{lamarre} Lamarre, J.-M., Puget, J.-L., Ade, P.~A.~R., et al.\ 2010, \aap, 520, A9 
		\bibitem[de Lapparent et al.(1986)]{lapparent} de Lapparent, V., Geller, M.~J., \& Huchra, J.~P.\ 1986, \apjl, 302, L1 
		\bibitem[Lee et al.(2015)]{lee} Lee, G.-H., Hwang, H.~S., Lee, M.~G., et al.\ 2015, \apj, 800, 80 
		\bibitem[Mart{\'{\i}}nez et al.(2016)]{martinez} Mart{\'{\i}}nez, H.~J., Muriel, H., \& Coenda, V.\ 2016, \mnras, 455, 127 
		\bibitem[Mennella et al.(2011)]{mennella} Mennella, A., Bersanelli, M., Butler, R.~C., et al.\ 2011, \aap, 536, A3 
		\bibitem[Murgia et al.(2010)]{murgia} Murgia, M., Govoni, F., Feretti, L., \& Giovannini, G.\ 2010, \aap, 509, A86 
		\bibitem[Nagai et al.(2007)]{nagai} Nagai, D., Vikhlinin, A., \& Kravtsov, A.~V.\ 2007, \apj, 655, 98
		\bibitem[Navarro et al.(1995)]{navarro} Navarro, J.~F., Frenk, C.~S., \& White, S.~D.~M.\ 1995, \mnras, 275, 56 
		\bibitem[Oegerle \& Hill(2001)]{oegerle} Oegerle, W.~R., \& Hill, J.~M.\ 2001, \aj, 122, 2858 
		\bibitem[Ostriker(1964)]{ostriker} Ostriker, J.\ 1964, \apj, 140, 1056 
		\bibitem[Piffaretti et al.(2011)]{mcxc} Piffaretti, R., Arnaud, M., Pratt, G.~W., Pointecouteau, E., \& Melin, J.-B.\ 2011, \aap, 534, A109 
		\bibitem[Planck Collaboration et al.(2011)]{planck1} Planck Collaboration, Ade, P.~A.~R., Aghanim, N., et al.\ 2011, \aap, 536, A1
		\bibitem[Planck Collaboration et al.(2013)]{planckgnfw} Planck Collaboration, Ade, P.~A.~R., Aghanim, N., et al.\ 2013, \aap, 550, A131
		\bibitem[Planck Collaboration et al.(2013)]{planck135} Planck Collaboration, Ade, P.~A.~R., Aghanim, N., et al.\ 2013, \aap, 550, A134
		\bibitem[Planck Collaboration et al.(2014)]{planckcosmo} Planck Collaboration, Ade, P.~A.~R., Aghanim, N., et al.\ 2014, \aap, 571, A16
		\bibitem[Planck Collaboration et al.(2014)]{pccs} Planck Collaboration, Ade, P.~A.~R., Aghanim, N., et al.\ 2014, \aap, 571, A28
		\bibitem[Planck Collaboration et al.(2014)]{mprox} Planck Collaboration, Ade, P.~A.~R., Aghanim, N., et al.\ 2014, \aap, 571, A29 
		\bibitem[Planck Collaboration et al.(2016)]{cosmo15} Planck Collaboration, Ade, P.~A.~R., Aghanim, N., et al.\ 2016, \aap, 594, A13 
		\bibitem[Planck Collaboration et al.(2016)]{szmap} Planck Collaboration, Aghanim, N., Arnaud, M., et al.\ 2016, \aap, 594, A22
		\bibitem[Planck HFI Core Team et al.(2011)]{phfi} Planck HFI Core Team, Ade, P.~A.~R., Aghanim, N., et al.\ 2011, \aap, 536, A4 
		\bibitem[Proust et al.(2006)]{shapley} Proust, D., Quintana, H., Carrasco, E.~R., et al.\ 2006, \aap, 447, 133 
		\bibitem[Rykoff et al.(2014)]{rykoff} Rykoff, E.~S., Rozo, E., Busha, M.~T., et al.\ 2014, \apj, 785, 104 
		\bibitem[Sakelliou \& Ponman(2004)]{sakelliou} Sakelliou, I., \& Ponman, T.~J.\ 2004, \mnras, 351, 1439
		\bibitem[Schaap(2007)]{dtfe1} Schaap, W.~E.\ 2007, Ph.D.~Thesis, 		
		\bibitem[Schaap \& van de Weygaert(2000)]{dtfe2} Schaap, W.~E., \& van de Weygaert, R.\ 2000, \aap, 363, L29 	
		\bibitem[Springel et al.(2005)]{springel} Springel, V., White, S.~D.~M., Jenkins, A., et al.\ 2005, Nature, 435, 629 
		\bibitem[Sunyaev \& Zeldovich(1969)]{sz} Sunyaev, R.~A., \& Zeldovich, Y.~B.\ 1969, \nat, 223, 721
		\bibitem[Tempel et al.(2014)]{tempel} Tempel, E., Stoica, R.~S., Mart{\'{\i}}nez, V.~J., et al.\ 2014, \mnras, 438, 3465 
		\bibitem[Teyssier et al.(2009)]{teyssier} Teyssier, R., Pires, S., Prunet, S., et al.\ 2009, \aap, 497, 335 
		\bibitem[Ulmer \& Cruddace(1981)]{ulmer} Ulmer, M.~P., \& Cruddace, R.~G.\ 1981, \apjl, 246, L99 
		\bibitem[Visvanathan \& Sandage(1977)]{visv} Visvanathan, N., \& Sandage, A.\ 1977, \apj, 216, 214 
		\bibitem[Vogelsberger et al.(2014)]{vol} Vogelsberger, M., Genel, S., Springel, V., et al.\ 2014, \mnras, 444, 1518 
		\bibitem[Wright et al.(2010)]{wise} Wright, E.~L., Eisenhardt, P.~R.~M., Mainzer, A.~K., et al.\ 2010, \aj, 140, 1868-1881
		\bibitem[York et al.(2000)]{sdss} York, D.~G., Adelman, J., Anderson, J.~E., Jr., et al.\ 2000, \aj, 120, 1579 
		\bibitem[Zhang et al.(2013)]{zhang} Zhang, Y., Dietrich, J.~P., McKay, T.~A., Sheldon, E.~S., \& Nguyen, A.~T.~Q.\ 2013, \apj, 773, 115 

\end{thebibliography}

\end{document}